\documentclass[aps,prl,twocolumn,superscriptaddress]{revtex4-1}

\usepackage[pdftex]{graphicx}

\begin{document}




\title{Millimeter-wave surface impedance of optimally - doped Ba(Fe$_{1-x}$Co$_{x}$)$_2$As$_2$ single crystals}

\author{A.~Barannik}
\affiliation{ Solid State Radiophysics Department, Institute of Radiophysics and Electronics, Nat. Acad.  Sci.  Ukraine, Kharkiv, Ukraine }

\author{N.~T.~Cherpak}
\email[Corresponding author: ]{ cherpak@ire.kharkov.ua }
\affiliation{ Solid State Radiophysics Department, Institute of Radiophysics and Electronics, Nat. Acad.  Sci.  Ukraine, Kharkiv, Ukraine }

\author{M.~A.~Tanatar}
\affiliation{The Ames Laboratory, Ames, Iowa 50011, USA}

\author{ S. Vitusevich}
\affiliation{ Institute of Bio- and Nanostructures, Forschungszentrum Juelich, Juelich, Germany  }

\author{ V. Skresanov}
\affiliation{ Solid State Radiophysics Department, Institute of Radiophysics and Electronics, Nat. Acad.  Sci.  Ukraine, Kharkiv, Ukraine }

\author{P.~C.~Canfield}

\affiliation{The Ames Laboratory, Ames, Iowa 50011, USA}
\affiliation{Department of Physics and Astronomy, Iowa State University, Ames, Iowa 50011, USA }

\author{R.~Prozorov}
\email[Corresponding author: ]{prozorov@ameslab.gov}
\affiliation{The Ames Laboratory, Ames, Iowa 50011, USA}
\affiliation{Department of Physics and Astronomy, Iowa State University, Ames, Iowa 50011, USA }

\date{24 December 2012}


\begin{abstract}

Precision measurements of active and reactive components of in-plane microwave surface impedance were performed in single crystals of optimally doped Fe-based superconductor Ba(Fe$_{1-x}$Co$_x$)$_2$As$_2$ ($x=$0.074, $T_c=$ 23 K). Measurements in a millimeter wavelength range ($K_a-$band, 35 to 40 GHz) were performed using whispering gallery mode excitations in the ultra-high quality factor quasi-optical sapphire disk resonator with YBa$_2$Cu$_2$O$_7$ superconducting ($T_c =$ 90 K) endplates.  The temperature variation of the London penetration depth is best described by a power law function, $\Delta \lambda (T) \sim T^n$, $n =$ 2.8, in a reasonable agreement with radio-frequency measurements on crystals of the same batch. This power-law dependence is characteristic of nodeless superconducting gap in the extended s-wave pairing scenario with a strong pairbreaking scattering. The quasiparticle conductivity of the samples, $\sigma_1 (T)$, gradually increases with the decrease of temperature, showing no increase below $T_c$, in a notable contrast with the behavior found in the cuprates. The temperature-dependent quasiparticle scattering rate was analyzed in a two-fluid model, assuming the validity of the Drude description of conductivity and generalized expression for the scattering rate. This analysis allows to estimate the range of the values of a residual surface resistance from 3 to 6 $m \Omega$.

\end{abstract}


\maketitle



\section{Introduction}

Determination of the superconducting gap structure plays important role in identification of the mechanism of superconductivity in recently discovered iron-arsenide superconductors \cite{Hosono}. That is why this problem was experimentally studied using a plethora of techniques, see e.g. reviews \cite{PaglioneNP,Johnston2010review,MazinNature,Hirschfeldreview2011,Prozorovreview2011,Stewart,Chubukovreview2012}.

Measurements of London penetration depth, $\lambda (T)$ provide an important insight into the temperature variation of the superfluid density, directly related to a superconducting gap structure. Several techniques were employed so far to study $\lambda (T)$ in iron pnictides. In particular, optimally - doped Ba(Fe$_{1-x}$Co$_x$)$_2$As$_2$ ($x \approx $0.07) has been studied by using several techniques that cover a wide range of frequencies. Single crystals were measured using essentially DC measurements by magnetic - force microscopy and scanning SQUID \cite{Luan2010,Luan2011FeCoL0}, radio - frequency tunnel - diode resonator \cite{GordonPRL,GordonPRB,irradiation} and muon spin rotation ($\mu$SR) in vortex state \cite{Williams2009,Williams2010FeCouSR} and in Meissner state \cite{uSR_MW_TDR_FeCo122_2012}, as well as microwave - range measurements \cite{Williams2009,Williams2010FeCouSR,Bobowski2010,uSR_MW_TDR_FeCo122_2012}. THz and optical refelectivity measurements were performed on thin films \cite{THz,optical1}.

Among these techniques, measurements of surface impedance allow determination of both active and reactive components of complex conductivity. This brings insight not only into the temperature-dependent London penetration depth, but also into the temperature-dependent quasi-particle scattering rate. Since anomalous scattering in the normal state is directly linked to the superconducting pairing strength \cite{Tailleferreview}, extension  of these measurements into a superconducting state is of notable interest. So far, only cavity perturbation technique has been used for microwave - range measurements \cite{Williams2009,Williams2010FeCouSR,Bobowski2010,uSR_MW_TDR_FeCo122_2012} and here we report the measurements using high $Q -$ factor quasi-optical resonator with high-$T_c$ superconducting end-plates. All these techniques consistently showed non - exponential power-law low - temperature behavior, $\Delta \lambda (T) =AT^n$, with $n \approx 2 - 2.8$ at the optimal doping in Ba(Fe$_{1-x}$Co$_x$)$_2$As$_2$ (``BaCo122'') \cite{GordonPRL,GordonPRB,irradiation,Williams2010FeCouSR,uSR_MW_TDR_FeCo122_2012,Luan2011FeCoL0}. Such behavior was ultimately attributed to the effect of strong pair - breaking scattering \cite{irradiation,Kogan2009,Gordonpairbreaking}, which actually supported $s_{\pm}$ pairing model \cite{irradiation}, although a possibility remains that nodes in predominantly c-axis direction may influence the in - plane penetration depth as well, since the latter is calculated by a full average over the Fermi surface \cite{Mishra2009}. Fully gapped superconductivity in BaCo122 at the optimal has been confirmed  by measurements of thermal conductivity \cite{TanatarPRL,ReidPRB}.

In this paper we report microwave surface impedance study of optimally - doped BaCo122 using novel and potentially very precise technique utilizing a high $Q -$ factor quasi-optical resonator with high-$T_c$ superconducting end-plates. Microwave surface impedance measurements allowed us to determine London penetration depth, which compares well with the results obtained on crystals from the same batch, providing a good reference point for our relatively novel measurement approach. We have also determined temperature - dependent  quasi-particle scattering time, which monotonically increases upon cooling below $T_c$.

\section{Experimental}

Single crystals of Ba(Fe$_{1-x}$Co$_x$)$_2$As$_2$ (BaCo122 in the following) were grown from FeAs:CoAs flux, as described in detail in Ref.~\onlinecite{NiNiCo}. Cobalt doping level, $x=0.074$, was determined using wavelength dispersive electron-probe spectroscopy (WDS). Superconducting transition temperature $T_c=$22.8~K, as determined in our microwave measurements, was in a good agreement with that determined on samples from the same batch in magnetization \cite{NiNiCo}, TDR \cite{GordonPRL} and resistivity \cite{Tanatar2009a} measurements.  For microwave surface impedance measurements samples were cleaved into a rectangular parallelepiped with dimensions $2.50 \times 3.50 \times 0.10$ mm$^3$.

\begin{figure}[tbh]%
\centering
\includegraphics[width=9cm]{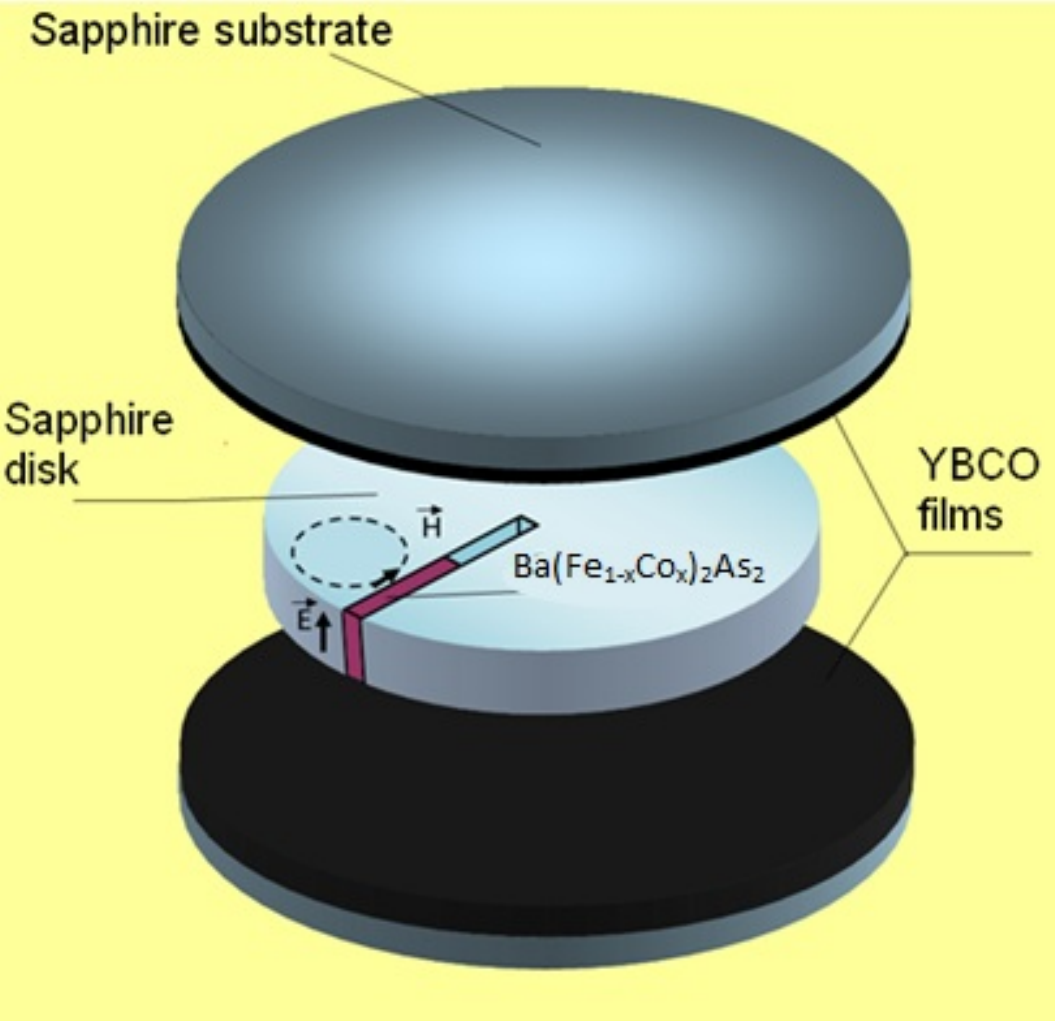}%
\caption{ Schematics of the slotted sapphire disk resonator experiment. The sapphire disk with single crystal of Ba(Fe$_{1-x}$Co$_x$)$_2$As$_2$ placed in the slot is sandwiched between superconducting YBa$_2$Cu$_3$O$_{7}$ film endplates. Whispering gallery mode excitation at a $K_a$ band frequency (35 to 40 GHz) produces an electric field $E$ parallel to the conducting plane of the sample, enabling measurements of in-plane surface impedance. }%
\label{f1_resonator}%
\end{figure}

Temperature-dependent microwave surface impedance, $Z_s=R_s+iX_s$ was measured  in $K_a$-band (35 - 40 GHz range) using sapphire disk quasi-optical resonator excited at whispering gallery modes (WGM). The resonator with conducting endplates (CEP) was developed earlier for the millimeter-wave impedance characterization of the cuprate high-$T_c$ films \cite{Cherpak20,Cherpak21}. For study of iron-pnictides it was modified into a disk resonator with a radial slot as illustrated in Fig.~\ref{f1_resonator}. This resonator geometry using CEPs made of YBa$_2$Cu$_3$O$_{7}$ (YBCO) films with $T_c \approx 90$~K was described in Ref.~\onlinecite{Barannik18}. It was developed specifically for precision measurements of microwave impedance properties of small-size superconductors with $T_c$ less than 90K.

\begin{figure}[tbh]%
\centering
\includegraphics[width=8cm]{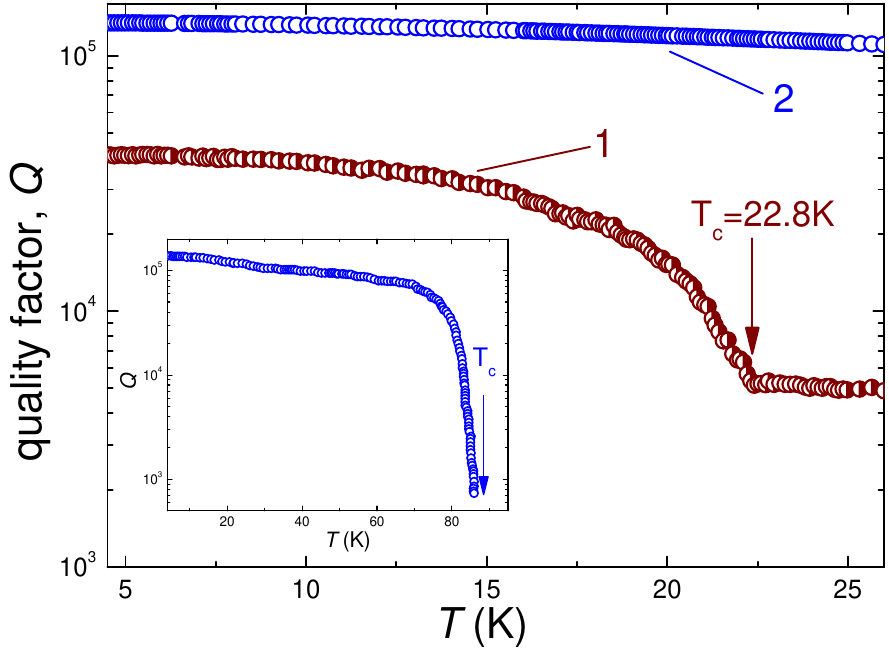} \\
\includegraphics[width=8cm]{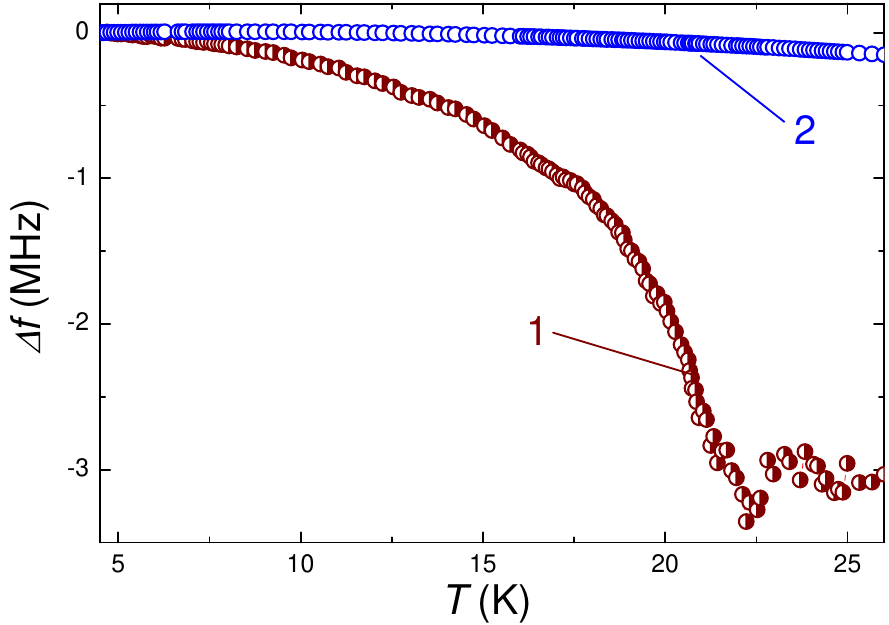}%
\caption{ Temperature-dependent (a) quality factor and (b) the resonant frequency shift of the resonator with single crystal Ba(Fe$_{1-x}$Co$_x$)$_2$As$_2$ (curves 1) and of the empty resonator (curves 2). Inset in panel (a) shows $Q(T)$ of the empty resonator in the temperature interval up to the superconducting transition of YBCO film endplates ($T_c$=90K). }%
\label{f2_Qf}%
\end{figure}

The resonator assembly, combining the sapphire disk excited at millimeter-wave WGMs and high-temperature superconducting CEP's, gives high quality factor, $Q \approx 10^5$, in temperature interval from 4.2~K up to about 30~K. The technique allows studying the microwave properties of unconventional superconductors in a range from millimeter to submillimeter wavelengths. For our measurements we used a novel technique for determining the frequency response of the resonators in the case of a partial removal of mode degeneration \cite{Barannik22}, as well as perturbed Lorenz form of the resonance line. Both modifications allowed us making precise determination of the resonance frequency and of the $Q-$factor \cite{ Skresanov23}, thus accurate measurements of the surface impedance. The measured $Q(T)$ and $f(T)$ for the empty resonator and the resonator with crystal under study are shown in Fig.~\ref{f2_Qf}.  The measurements of $Q$ factor were performed with the weak coupling of the resonator with the dielectric waveguides. The coupling is limited by the sensitivity of the measuring apparatus (HP8510C vector network analyzer). The obtained value of $Q$ can be taken as the intrinsic $Q$-factor with high accuracy. The accuracy of the determined resonance frequency depends on the $Q$-factor. In our case the accuracy of the resonant frequency is about few kHz in the K band. The decreased accuracy of frequency shift measurement at $T>T_c$ is due to a considerable decrease of the resonator $Q-$factor when the sample becomes normal above $T_c$.

\section{Analysis of surface impedance}

To obtain $R_s(T)$ and $X_s(T)$ from measured $Q(T)$ and $f(T)$ we used known expressions (see  Refs.~\onlinecite{Bonn24,Cherpak20,Cherpak21}). The surface resistance $R_s(T)$ of the sample can be determined from the variation of the $Q-$factor of the resonator as:

\begin{equation}
\label{1}
 A_s R_s(T)=\Delta Q^{-1}(T)
\end{equation}

\noindent
Since it is impossible to determine accurately the eigenvalue of the frequency of the resonator with perfectly conducting elements (this obstacle is general for all types of the resonators), one can obtain the expression for the temperature variation of the surface reactance $\Delta X_s(T)$ through the temperature change of the resonator frequency $\Delta\omega(T)$ as:

\begin{equation}
\label{2}
A_s \Delta X_s(T)=-2\Delta \omega(T)/\omega (T),
\end{equation}

\noindent
where $\omega=2\pi f$, $A_s$ is the filling factor that depends on the geometry and dimensions of the sample as well as on the field distribution (mode) in the resonator. The coefficient $ A_s$ can be calculated analytically by solving the resonator electromagnetics problem \cite{Cherpak20}. If the analytical solution cannot be found, the value of $A_s$ can be determined by a calibration procedure, using samples with known properties \cite{Cherpak21}. In this work we evaluated $A_s$ by simulating the resonator response by using CST Microwave Studio program assuming perfect dielectric loss ($\tan \delta$=0), perfect CEP ($R_s^{CEP}=0$, see below) and a conductor with the preselected surface resistance $R_s^e(T)$ with the dimensions identical to those of our sample. In this case the following equation can be used:

\begin{equation}
\label{3}
(Q_{os}^e)^{-1}= A_s R_s^e
\end{equation}

\noindent
Since the CST Microwave Studio program does not account for radiation losses, the calculated value of $(Q_{os}^e)^{-1}$ for the preselected $R_s^e$ and the mode in the resonator gives the undetermined constant $A_s=1/ R_s^e Q_{os}^e$. For $R_s^e$=50m$\Omega$ we obtained $Q_{os}^e$=70672 and $A_s=2.83 \times 10^{-4}$ m$\Omega ^{-1}$ for the interaction of HE$_{1610}$-mode with a sample of $2.50 \times 3.50 \times 0.10$ mm$^3$.

In our case of the open dielectric resonator, instead of using Eq.~(1) it is necessary to use more general approach taking into account additive character of microwave losses in the resonator with the sample under test and without it:

\begin{equation}
\label{4}
   Q_0^{-1}=k \tan \delta +A_s^{CEP}R_s^{CEP}+Q_{rad 0}^{-1}
\end{equation}

\begin{equation}
\label{5}
Q_{0S}^{-1}=k \tan \delta +A_s^{CEP}R_s^{CEP}+ A_sR_s + Q_{rad S}^{-1}
\end{equation}

\noindent
Here $A_s^{CEP}$ and $R_s^{CEP}(T)$ are the filling factor and surface resistance values of CEP, $k$ is the coefficient very close to 1 \cite{Cherpak20}, $\tan \delta$ is the loss tangent of the sapphire dielectric, $Q_{rad}$ is the $Q$-factor determined by the radiation losses. Unlike the case of the homogeneous sapphire disk QDR, where CEP $Q_{rad,0}^{-1}<10^{-9}$ and radiation loss can be neglected \cite{Cherpak20}, in the radially slotted QDR a value of $Q_{rad}^{-1}$ becomes comparable with other losses in Eqs.(4) and (5). In addition, the values of $Q_{rad 0}^{-1}$  and $Q_{rad S}^{-1}$ are different and cannot be determined with suitable accuracy, which does not allow finding $R_s$ directly from Eqs.(4) and (5). However taking into account temperature independence of $Q_{rad}^{-1}$, one can find the temperature difference $ \Delta R_s(T)$ in comparison with $R_s$ at a certain reference temperature $T_{ref}$. In this case, instead of Eq.(3), we can obtain a simpler expression:

   \begin{equation}\label{6}
\Delta R_s(T,T_{ref})=\frac{\Delta Q_{0S}^{-1}(T,T_{ref})- \Delta Q_0^{-1}(T,T_{ref})}{A_s},
   \end{equation}

\noindent
here $\Delta Q_{0S}^{-1}(T,T_{ref})= Q_{0S}^{-1}(T)- Q_{0S}^{-1}(T_{ref})$,
and $\Delta Q_{0}^{-1}(T,T_{ref})= Q_{0}^{-1}(T)- Q_{0}^{-1}(T_{ref})$.
As a rule, $T_{ref}$ is the lowest available temperature. Evidently $\Delta R_s(T,T_{ref}) =R_s(T)-R_s(T_{ref})$ and
$\Delta R_s(T>T_c,T_{ref}) =R_s(T>T_c)-R_s(T_{ref})$. Because in the normal state $R_s(T>T_c)=X_s(T>T_c)$, we can write $R_s(T_{ref})=X_s(T>T_c)-\Delta R_s(T>T_c,T_{ref})$. Thus we have $R_s^{\prime}(T)=R_s(T_{ref}) +\Delta R_s(T,T_{ref})$. The measured temperature dependence $\Delta R_s(T,T_{ref})$ allows us to extrapolate $R_s^{\prime} (T)$ to $R_s(T \rightarrow 0)=R_{res}$ and obtain the whole temperature dependence,

\begin{equation}\label{7}
R_s(T)=R_{res}+ \Delta R_s(T)
\label{7}
\end{equation}

\noindent
where $R_{res}$ is residual resistance, which has certain value but difficult to assign because of its small magnitude.

Surface reactance $X_s(T)$ is also an important characteristic of the sample. However, it is difficult to obtain the absolute value of $ X_s(T)$, with the main problems coming from the impossibility to determine the eigenvalue frequency of the resonators with perfect conducting surfaces, as mentioned above, and insufficient reproducibility of the frequencies upon reassembling the resonator. Evidently, in our case of the radially slotted QDR, similar to other resonator techniques, the most appropriate approach is to determine reactance variation $X_s(T)$ using the following relation \cite{Cherpak21}

   \begin{equation}\label{8}
\Delta X_s(T,T_{ref})=X_s(T,T_{ref})-X_s(T_{ref})
   \end{equation}

\noindent
Because $X_s(T)= \omega (T) \mu_0 \lambda (T)$, where $\lambda (T)$ is London penetration depth, we can write
   \begin{equation}\label{9}
\Delta X_s(T,T_{ref})=\omega (T) \mu _0 \Delta \lambda (T,T_{ref}),
   \end{equation}

\noindent
where
   \begin{equation}\label{10}
\Delta \lambda (T,T_{ref})= \lambda (T) - \lambda (T_{ref})
   \end{equation}

\noindent
From (2) and (9) $\Delta \lambda (T, T_{ref})$ can be expressed as
   \begin{equation}\label{11}
\Delta \lambda (T, T_{ref}) = -\frac{2 \Delta \omega (T,T_{ref})}{A_s \omega ^2 (T) \mu_0}
   \end{equation}

\noindent
where $\Delta \omega (T,T_{ref})= \omega (T)- \omega (T_{ref})$.
Using $\lambda (0)$ determined from other measurements and the experimental $\Delta \lambda (T, T_{ref})$ extrapolated to
$T \rightarrow 0$, $\lambda (T)$ can be calculated as: $\lambda (T) = \lambda (0)+ \Delta \lambda (T_{ref}, 0)+\Delta \lambda (T, T_{ref})$. Usually $\Delta \lambda (T_{ref},0) \ll \lambda (0)$, therefore the error in finding this value does not influence noticeably the accuracy of $X_s(T)$ determination

\begin{equation}
\label{12}
X_s (T) = \omega \mu _0 [\lambda (0) +\Delta \lambda (T_{ref},0)]-\frac{2 \Delta \omega (T, T_{ref})}{A_s \omega (T)}
\end{equation}

\noindent
It should be noted that in $\Delta \omega (T, T_{ref}$), the variations of $\Delta \omega _{\epsilon} (T, T_{ref}$)  and of $\Delta \omega_d (T, T_{ref}$) determined by the temperature dependence of both of sapphire permittivity, $\epsilon$ , and of the disk dimensions, are removed by subtracting the $ f(T)=\omega/2\pi$ curve from the curve obtained from the experimental data (see Fig.~\ref{f2_Qf}b).

   Using known values of $R_s$ and $X_s$ at $\omega \tau \ll1$, where $\tau$ is the quasiparticle scattering time, one can find conductivities $\sigma_1$ and $\sigma_2$ from $Z_s=[i\omega\mu _0/(\sigma_1-i\sigma_2)]^{1/2}=R_s+iX_s$, as (see Ref.~\onlinecite{Hensen25})

\begin{equation}
\label{13}
\sigma _1=2 \omega \mu _0 \frac{R_s X_s}{(R_s^2+X_s^2)^2}
\end{equation}

\begin{equation}
\label{14}
\sigma _2= \omega \mu _0 \frac{(X_s^2-R_s^2)}{(R_s^2+X_s^2)^2}
\end{equation}

\noindent
where $\sigma_1=\sigma _n$ is the real part of the quasiparticle conductivity in a microwave range (it represents the loss related to the conductivity of the normal carriers – quasiparticles), and $\sigma_2=1/(\omega\mu _0 \lambda ^2)=-i\sigma_s$ represents the kinetic energy of the superconducting carriers. Assuming that both Drude formula for conductivity, $\sigma _n= \frac{e^2 n_n \tau}{m}\frac{1}{(1+I \omega \tau)}$, at $\omega \tau \ll 1$, and the equation $n_s(0)-n_s(T)=n_n(T)$ for two-fluid model are valid, we can obtain the expression for the quasiparticle scattering rate in a form of

\begin{equation}
\label{15}
\tau^{-1} (T)= \frac{1-\frac{\lambda (0) ^2}{\lambda (T) ^2}}{\mu_0 \sigma _1(T) \lambda (0) ^2}
\end{equation}

In a more general case of arbitrary $\tau$, the quasiparticle conductivity also becomes a complex number, $\sigma_1=\sigma^{\prime}_1-i\sigma^{\prime \prime}_1$, where $\sigma^{\prime \prime}_1=\omega \tau \sigma^{\prime}_1$. In this case we have to replace the $\sigma_1$  by  $\sigma^{\prime}_1$ and the $\sigma_2$  by  $\sigma_2+\sigma^{\prime \prime}_1$ in equations (13) and (14). It should be emphasized that only $\sigma^{\prime}_1$ and $\sigma_2+\sigma^{\prime \prime}_1$  are determined based on the experimental values $R_s$ and $X_s$. Then the ratio of values $\sigma_2$ and $\sigma^{\prime}_1$  can be obtained:

\begin{equation}
\label{16}
\sigma_2/\sigma_1^{\prime}=\frac{(X_s^2-R_s^2)}{2X_sR_s}- \omega \tau
\end{equation}

\noindent
and on the other hand the following expression can be obtained:

\begin{equation}
\label{17}
\sigma_2/\sigma_1^{\prime}=\frac{n_s}{n_n}\frac{[1+( \omega \tau)^2]}{\omega \tau}
\end{equation}

Using the condition, $n_s(0)-n_s(T)=n_n(T)$, and expressions (16) and (17) we derive the following expression:

\begin{equation}
\label{18}
\frac{1+(\omega \tau )^2}{\omega \tau}=\frac{1}{\frac{\lambda _L^2(T)}{\lambda_L^2(0)}-1}(\frac{X_s^2-R_s^2}{2X_sR_s}-\omega \tau)
\end{equation}

The square of the London penetration depth, $\lambda^2(T)$, can be rewritten in terms of $\sigma_2(T$) as $ \lambda^2 (T)= \frac{1}{\omega \mu _0 \sigma _2 (T)}$ and further in terms of $R_s$ and $X_s$ using expression (16):

   \begin{equation}\label{19}
\lambda _L^2(T)=\frac{1}{\omega \mu_0 \sigma _1^{\prime}}\frac{1}{\frac{X_s^2-R_s^2}{2X_sR_s}-\omega \tau}
   \end{equation}

Equations (18) and (19) are now used to obtain a relation for the scattering rate

   \begin{equation}\label{20}
\tau ^{-1} (T)= \frac{1}{\mu_0 \lambda^2(0) \sigma_1 ^{\prime}}-\frac{X_s^2-R_s^2}{2\omega X_s R_s},
   \end{equation}

\noindent
connecting it with the measured experimental quantities, $R_s$ and $X_s$. Equation (20) is true for arbitrary correlation of $R_s(T)$ and $X_s(T)$. When $R_s(T) \ll X_s(T)$, or $\omega \tau \ll 1$, the relation (20) can be reduced to (15).

\section{Results and discussion}

\begin{figure}[tbh]%
\centering
\includegraphics[width=9cm]{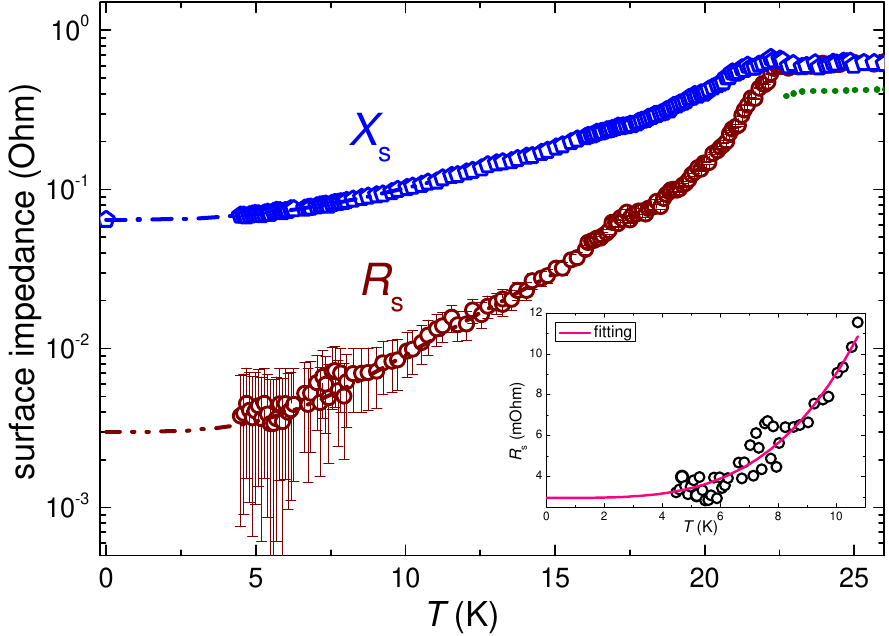}%
\caption{ Temperature-dependent surface impedance of the single crystal of optimally - doped Ba(Fe$_{1-x}$Co$_x$)$_2$As$_2$, $x=0.074$. Inset shows $R_s(T)$ in the low temperature range.}%
\label{f3_RsXs}%
\end{figure}

The experimentally determined $R_s(T)$ and $X_s(T)$ are shown in Fig.~\ref{f3_RsXs}. The value of $X_s(T)$ in the normal state was determined from the measured $\Delta X_s (T)$ and calibrated using the value of the penetration depth $\lambda (0)$=210~nm determined by the tunnel-diode resonator technique \cite{Gordonlambda0}. The $R_s(T)$ was determined by measuring $Q$-factors of the resonator with the sample and without it and using $R_s(T)=X_s(T)$ at $T\geq T_c$. The residual surface resistance (per square, so it is measured in Ohm), $R_{res} \approx 3 - 6$ $m\Omega$ was estimated from $T \rightarrow 0$ extrapolation of $R_s(T)$ (see inset in Fig.~\ref{f3_RsXs}). The accuracy of thus determined $R_{res}$ depends on the accuracy of both $X_s(T)$ at low temperatures and of $R_s(T)$ and $X_s(T)$ at $T>T_c$. The precise determination of $R_{res}$ is especially important for unconventional superconductors, because often the values of $R_{res}$ in these materials are much higher than in conventional BCS superconductors \cite{Hensen25} and the nature of this phenomenon is so far not understood \cite{Hein1999,Barannik2008}.
A value of $R_s(T>T_c)$ can also be found using experimental measurement of the sample resistivity $\rho$ as $R_s=(\omega \mu \rho/2)^{1/2}$. Thus determined values are shown in Fig.~\ref{f3_RsXs} with the dotted line. One can see that these values are a slightly smaller than $R_s(T>T_c)$ obtained from the calibration using $\lambda (0)$=210 nm. This discrepancy can be explained by the roughness of the sample surface, because $R_s$ can only increase compared to an ideally smooth surface, or by adding some anomalous character to the normal skin-effect.

\begin{figure}[tbh]%
\centering
\includegraphics[width=8cm]{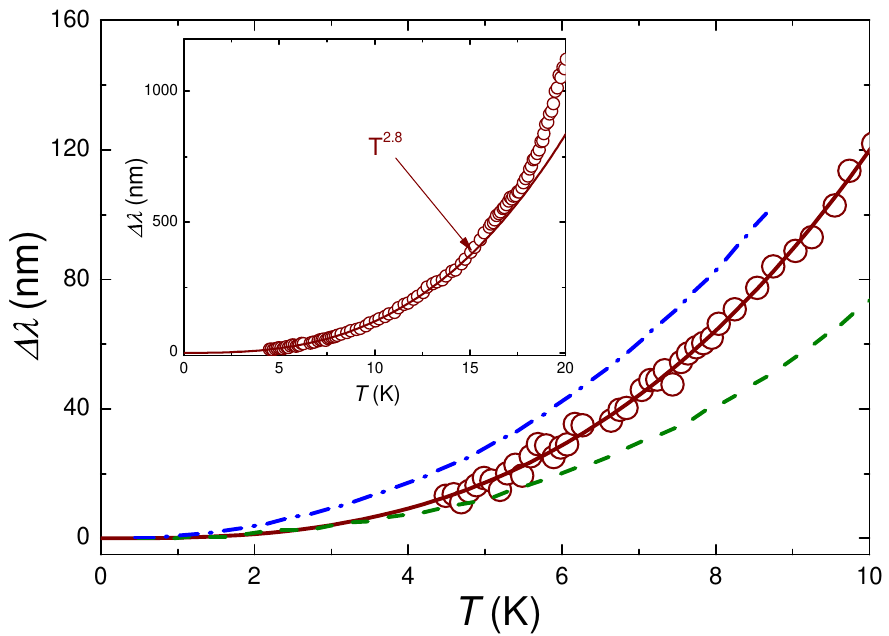}%
\caption{ The London penetration depth $\lambda (T)$ in single crystal of optimally - doped Ba(Fe$_{1-x}$Co$_x$)$_2$As$_2$, $x=0.074$, at $T<T_c/2$. The solid line is the power - law fit, $\sim T^{2.8}$, the dashed and the dot-and-dashed lines correspond to the experimental data of Refs.~\onlinecite{GordonPRL} and \onlinecite{GordonPRB}, respectively. The open circles represent experimental data. Inset shows $\lambda (T)$ in a broader temperature range. }%
\label{f4_lambda}%
\end{figure}

\begin{figure}[tbh]%
\centering
\includegraphics[width=8cm]{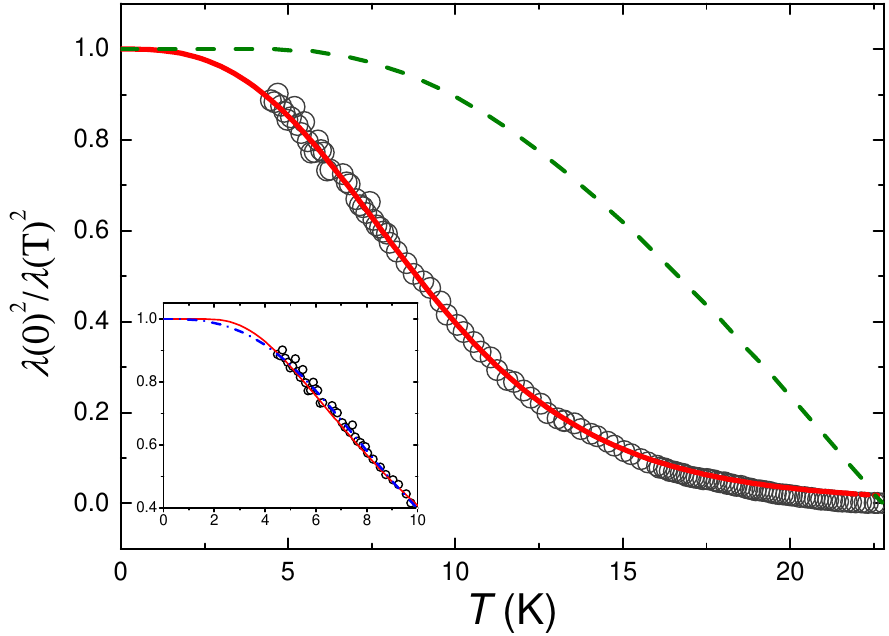}%
\caption{ The temperature-dependent superfluid density in a single crystal of optimally - doped Ba(Fe$_{1-x}$Co$_x$)$_2$As$_2$, $x=0.074$. The solid lines correspond to the power - law $\Delta \lambda \sim T^{2.8}$. A dashed line corresponds to a single - gap isotropic s-wave BCS superconductor with $\Delta=1.76k_BT_c$. Inset shows $n_s(T)$ calculated with $\Delta \lambda (T) \sim T^{2.8}$ (solid line) and the best exponential fit resulted in $\Delta=0.75k_BT_c$ (dash-dotted line).}%
\label{f5_superfluid}%
\end{figure}

The London penetration depth $\lambda (T)$ determined at low temperatures, $T<T_c/2$, from microwave data is shown in Fig.~\ref{f4_lambda}. The observed temperature variation is best described by the power-law, $\Delta \lambda (T) \sim T^n$ with $n =$ 2.8 from low temperatures up to at least $0.6T_c$. This dependence is similar to the one obtained in the radio - frequency TDR measurements \cite{Prozorovreview2011}, especially on the high-quality crystals \cite{irradiation}. A similar exponent was determined in another microwave impedance study performed by cavity perturbation technique at 13 GHz, indicating the exponent of $n =$2.66 \cite{Bobowski2010}. The variation, $\Delta \lambda (T)=\lambda (T)-\lambda (0)$ and the full superfluid density, $n_s(T) =\left[\lambda (0)/\lambda (T)\right]^2$  are commonly used to analyze the penetration depth data and compare the results with calculations for various superconducting gap structures \cite{Prozorovreview2011}. In Fig.~\ref{f5_superfluid}, temperature - dependent $n_s(T)$ was constructed from $\lambda (T)$ determined from $\sigma_2(T)$ under the condition of $\omega \tau \ll 1$ in eq.~(14). Solid line shows a power - law fit, corresponding to $\Delta \lambda (T) \sim T^{2.8}$ and the dashed line shows expectation for isotropic weak coupling single - gap s-wave BCS superconductor with $\Delta=1.76k_BT_c$. Inset in Fig.~\ref{f5_superfluid} compares calculated $n_s(T)$ with $\Delta \lambda (T) \sim T^{2.8}$ and for the exponential variation with $\Delta=0.75k_BT_c$ obtained from the best exponential fit at the low temperature interval. Clearly, power-law behavior with $n=$2.8 provides the best description of the data. However in the low temperature interval it is impossible to say what temperature dependence gives better fitting. The fact that lowest temperatures can also be described by the exponential fit with smaller than weak - coupling $1.76k_BT_c$ value of $0.75k_BT_c$ simply means that we are dealing with a two - gap system. The convex shape of $n_s(T)$ at the elevated temperatures supports the multi - gap behavior \cite{Prozorovreview2011}.

\begin{figure}[tbh]%
\centering
\includegraphics[width=8cm]{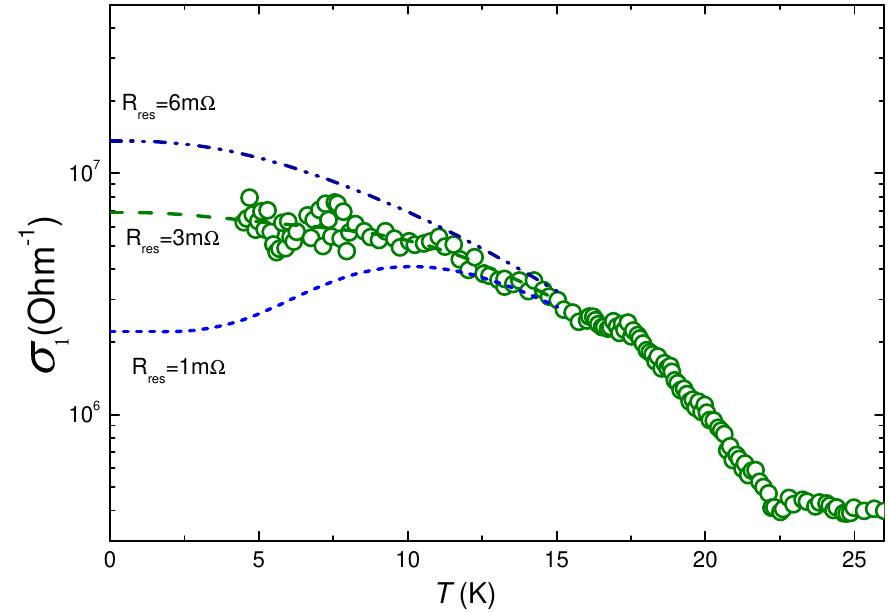}%
\caption{Temperature-dependent quasiparticle conductivity, $\sigma _1 (T)$, in a single crystal of optimally - doped Ba(Fe$_{1-x}$Co$_x$)$_2$As$_2$, $x=0.074$, calculated for different values of $R_{res}$.
}%
\label{f6_sigma1}%
\end{figure}

Figure \ref{f6_sigma1} shows the temperature-dependent quasiparticle conductivity $\sigma_1$, calculated using eq. (13). The quasi-particle conductivity $\sigma_1(T)$ increases on cooling, - a behavior similar to that found previously in YBCO cuprate superconductor \cite{Bonn31} and recently in Fe-based pnictides PrFeAsO$_{1-y}$ \cite{microwavePr1111}, Ba$_{1-x}$K$_x$Fe$_2$As$_2$ \cite{microwaveBaK} and FeSe$_{0.4}$Te$_{0.6}$ \cite{Takahash2011MWFeSeTe}. However the accurate value of $\sigma_1(T)$ in our measurements depends strongly on the correct determination of  the residual surface resistance $R_{res}$ (see discussion above) and needs further studies. The observed $ \sigma_1(T)$ can be explained by a strong temperature dependent quasiparticle scattering rate decreasing rapidly with temperature. Similar tendency was found in FeSe$_{0.4}$Te$_{0.6}$ \cite{Takahash2011MWFeSeTe} where the Authors propose a crossover from dirty at $T_c$ to clean limit at the low temperatures to explain the convex shape of $n_s(T)$ and this idea requires further investigation.

Another important feature of $\sigma_1(T)$ (Fig.~\ref{f6_sigma1}) is the absence of a peak below or at $T_c$. $\sigma_1(T)$ changes monotonically through $T_c$, similar to previous microwave  measurements, Ref.~\onlinecite{Bobowski2010}. This contradicts the results obtained from measurements on thin films in terahertz \cite{THz} and optical \cite{optical1} domains. This may point to significant difference between high - quality single crystals and thin films of Fe - based superconductors.

\begin{figure}[tbh]%
\centering
\includegraphics[width=8cm]{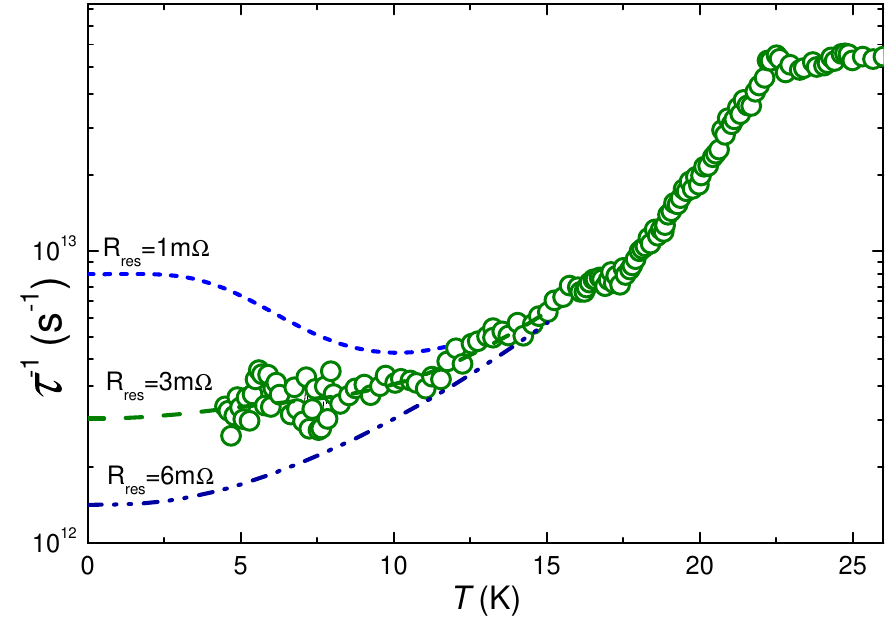}%
\caption{ The temperature-dependent quasiparticle scattering rate, $\tau ^{-1}$, in single crystal of optimally - doped Ba(Fe$_{1-x}$Co$_x$)$_2$As$_2$, $x=0.074$, calculated for different values of $R_{res}$.
}%
\label{f7_tau}%
\end{figure}

Figure ~\ref{f7_tau} shows that the temperature dependence of $\tau^{-1}(T)$ is determined  by the quasiparticle conductivity. This suggests that inelastic scattering plays an important role in iron pnictide superconductors even at very low temperatures deep into the superconducting state. It also follows from Fig.~\ref{f7_tau} that the  selection of $R_{res}$=1 $m\Omega$ gives unphysical result of the scattering rate that would increase as $T \rightarrow 0$. This allows us to narrow the range of $R_{res}$ from 3 to 6 $m\Omega$. The strong temperature dependence of $\tau ^{-1}(T)$ was also observed in the cuprates \cite{Bonn24} as well as in other pnictides, FeSe$_{0.4}$Te$_{0.6}$ \cite{Takahash2011MWFeSeTe} and Ba$_{1-x}$K$_x$Fe$_2$As$_2$ \cite{Schachinger}. It seems to be a general feature for all of the unconventional superconductors.

\section{Conclusion}

In conclusion, microwave (35 to 40 GHz) measurements of the in-plane London penetration depth and quasiparticle conductivity using high quality - factor quasi - optical resonator with high-$T_c$ superconducting end-plates were performed on single crystals of optimally - doped Ba(Fe$_{1-x}$Co$_x$)$_2$As$_2$, $x=0.074$, $T_c$=22.8~K. The London penetration depth varies as a power-law, $\Delta \lambda (T)= AT^n$ with the exponent n=2.8, consistent with previous studies. The temperature-dependent quasiparticle conductivity, $\sigma _1 (T)$ does not show a peak below or at $T_c$, consistent with another microwave study at the lower frequency of 13 GHz \cite{Bobowski2010}, but in a stark disagreement with optical and THz measurements on thin films \cite{THz,optical1} and with rather low frequency measurements at 50 kHz \cite{Yong2011}. The quasiparticle conductivity increases monotonically on cooling below $T_c$ suggesting strong inelastic scattering even at low temperatures.

\section{Acknowledgements}

We thank Prof. N.~Klein for support and collaboration at Forschungszentrum Juelich, Juelich, Germany. Work at the Ames Laboratory was supported by the Division of Materials Science and Engineering, Basic Energy Sciences, Department of Energy (USDOE), under Contract No. DEAC02-07CH11358. Work in Kharkiv was supported by Department of Radiophysics and Electronics, IRE NAS of Ukraine under Project State No. 0106U011978.


\end{document}